\documentclass[aps, 10pt, prd,
               notitlepage, twocolumn, superscriptaddress,
               nofootinbib, floatfix,
               longbibliography]{revtex4-1}

\usepackage{amsmath}
\usepackage{amssymb}
\usepackage{amsfonts}
\usepackage[utf8]{inputenc}
\usepackage[T1]{fontenc}
\usepackage{mathrsfs}
\usepackage{anyfontsize}
\usepackage{aas_macros}

\usepackage[linktocpage,breaklinks]{hyperref}
\usepackage[usenames,dvipsnames]{xcolor}

\usepackage{txfonts}
\usepackage{tensor}
\usepackage{bm,float}

\usepackage{graphicx}
\usepackage{epsfig}
\usepackage{epstopdf}
\usepackage{dsfont}

\usepackage{natbib}
\usepackage{hyperref}
\usepackage{verbatim}

\hypersetup{colorlinks=true,
            citecolor=MidnightBlue,
            linkcolor=MidnightBlue,
            urlcolor=MidnightBlue}

\AtBeginDocument{%
    \newwrite\bibnotes
    \def\bibnotesext{Notes.bib}
    \immediate\openout\bibnotes=\jobname\bibnotesext
    \immediate\write\bibnotes{@CONTROL{REVTEX41Control}}
    \immediate\write\bibnotes{@CONTROL{%
    apsrev41Control,author="08",editor="1",pages="1",title="0",year="1"}}
    \if@filesw
    \immediate\write\@auxout{\string\citation{apsrev41Control}}%
    \fi
}

\begin{document}

\title{Equation-of-State Constraints on the Neutron-Star Binding Energy \\ and Tests of Binary Formation Scenarios}

\author{A. Miguel Holgado}
\affiliation{McWilliams Center for Cosmology and Department of Physics, Carnegie Mellon University, Pittsburgh, Pennsylvania, 15213, USA}
\email{mholgado@andrew.cmu.edu}

\date{\today}

\begin{abstract}
The second supernova that forms double-neutron-star systems is expected to occur in a progenitor that is ultra-stripped due to binary interactions.
Thus, the secondary neutron star's mass as well as the post-supernova binary's orbital parameters will depend on the nature of the collapsing progenitor core. 
Since neutron stars are in the strong-gravity regime, their binding energy makes up a significant fraction of their total mass-energy budget.
The second neutron star's binding energy may thus provide a unique insight as to whether its progenitor was a low-mass iron core or an oxygen-neon-magnesium core. 
I obtain relations for the baryonic mass and binding energy incorporating both a hadronic equation-of-state catalog as well as recent multi-messenger neutron-star observations.  
With these relations, I obtain the first tight constraints on the baryonic mass and binding energy of three neutron stars that are thought to have formed from an ultra-stripped progenitor.
With these constraints, I test if each neutron star is consistent with forming from an ONeMg core that undergoes an electron-capture supernova.
From these tests, I find that this scenario can be ruled out for one of three neutron stars. 
Neutron-star properties and the dense-matter equation of state can thus help distinguish binary formation scenarios. 
\end{abstract}

\date{{\today}}

\maketitle

\textbf{Introduction}.
In order for isolated double-neutron-star (DNS) systems to merge within the age of the Universe, binaries need to survive two supernova explosions and one or more common-envelope phases \citep{tauris_formation_2017}. 
The second supernova is expected to occur in an ultra-stripped progenitor, whose outer envelope has been removed due to binary interactions.
The collapsing core of such a star may either be an iron core or an oxygen-neon-magnesium (ONeMg) core \citep{tauris_ultra-stripped_2015,poelarends_electron_2017}.
Transient surveys have been finding candidate ultra-stripped supernovae that likely lead to DNS formation, thus providing new constraints with which to test explosion models \citep{de_hot_2018,yao_sn2019dge_2020,nakaoka_calcium-rich_2020,hiramatsu_electron-capture_2020}. 
\par
NSs are in the strong-gravity regime, so a sizable fraction of their mass-energy budget is in gravitational binding energy. 
How much binding energy a NS has depends on the equation of state (EoS), where EoSs that produce more compact stars will have larger binding energies, and vice versa. 
Astronomical observations of NSs have provided independent measurements on various NS properties, including the tidal deformability (from gravitational waves \citep{ligo_scientific_collaboration_and_virgo_collaboration_gw170817:_2017,ligo_scientific_collaboration_and_virgo_collaboration_properties_2019}), the mass and radius (from high-cadence X-ray timing \citep{riley_nicer_2019,miller_psr_2019,raaijmakers_nicer_2019}), as well as lower limits to the maximum mass (from radio timing \citep{cromartie_relativistic_2020}). 
All of these multi-messenger observations have provided tighter and tighter constraints on the dense-matter EoS. 
\par
These constraints can also be used to characterize the NS binding energy and how it may play a role in DNS formation \citep{holgado_role_2021}. 
A NS's binding energy and thus its baryonic mass may also provide independent clues regarding the core of its progenitor star \citep{newton_testing_2018}. 
Here, I consider three NSs that have precisely measured masses and that are candidates for forming from an ultra-stripped progenitor: pulsar B of the double pulsar J0737-3039A/B \citep{burgay_increased_2003,lyne_double-pulsar_2004} (J0737B, herein), and the companions to J1829+2456 \citep{haniewicz_precise_2021} and J1756-2251 \citep{ferdman_psr_2014} (J1829c and J1756c, respectively, herein)\footnote{The systems J0737-3039A/B as well as J1756-2251 and its companion were discovered with the Parkes Observatory, while J1829+2456 and its companion were discovered with the Arecibo Observatory.}.
I then test whether or not each pulsar may have formed from an iron core or an ONeMg core. 
\par
\textbf{Iron Cores vs.~ONeMg Cores.}
The mass loss and NS kick from an ultra-stripped supernova \citep{beniamini_formation_2016,abbott_progenitor_2017,shao_role_2018,giacobbo_impact_2019,holgado_gravitational_2019} depends on the secondary's progenitor core composition. 
Iron cores can span a wide range of masses that lead to a wide range of remnant NS gravitational masses ${\simeq} 1.1-1.8 M_\odot$, where the progenitor core mass loss is $\sim 0.1 M_\odot$ and the kicks can be large $\gtrsim 100$ km/s \citep{tauris_ultra-stripped_2015}.
However, recent multi-dimensional simulations of ultra-stripped supernovae with low-mass iron cores suggest that the kicks can be low, i.e., 10s of km/s, depending on the structure of the progenitor star \citep{muller_multidimensional_2018,muller_three-dimensional_2019}.  
\par
For ONeMg cores, runaway electron captures onto Ne and Mg occur when a critical density and temperature are reached in the core \citep{miyaji_supernova_1980,nomoto_evolution_1984,miyaji_collapse_1987}.
These electron captures reduce the pressure that opposes gravity, prompting a core collapse that results in an electron-capture supernova (ECSN). 
The corresponding ONeMg core mass range for ECSN is expected to be fairly small ($\approx 1.36-1.38 M_\odot$), so the range of remnant NS gravitational masses is narrow. 
Just before collapse, the progenitor's density profile steeply decreases beyond the ONeMg core, so the explosion occurs before momentum significantly builds up on the remnant NS. 
The kicks from ECSNe are thus expected to be 10s of km/s at most. 
\par
The core mass loss is expected to be ${\lesssim}10^{-2} M_\odot$, such that the progenitor ONeMg core mass can be considered to be a reasonable estimate of the remnant NS's baryonic mass \citep{podsiadlowski_double_2005}. 
Here, I use the ONeMg core mass range $1.357-1.377 M_\odot$ \citep{takahashi_evolution_2013} as the range for which ECSNe occur\footnote{Binary stellar evolution models may estimate a larger upper limit on the baryonic mass of $\le 1.43 M_\odot$ for ECSNe \citep{tauris_ultra-stripped_2015}.
However, direct numerical modeling of a possible ONeMg core collapse leading to an ECSN has not yet been carried out at these higher masses.}. 
If a NS's baryonic mass falls within this range, allowing for some finite core mass loss, then the NS could have formed via an ECSN.
If a NS's baryonic mass is significantly outside of this range, either the NS formed from an iron core collapse or our current understanding of ECSNe needs to be revised. 
\par
\textbf{The Binding Energy and Baryonic Mass.} 
The binding energy and baryonic mass of a NS have as of yet neither been directly measured nor observed. 
However, one can rely on other macroscopic NS properties that have been measured to constrain what the binding energy and baryonic mass should be. 
Properties such as the moment of inertia and tidal deformability have been shown to obey relations that are insensitive to the EoS \citep{yagi_i-love-q_2013,yagi_i-love-q_2013-1}. 
Relations between the binding energy and the mass or compactness have also been obtained (with larger scatter) \citep{lattimer_analysis_1989,lattimer_neutron_2001,breu_maximum_2016}, which can then be used to constrain the baryonic mass when given a gravitational mass measurement. 
\par
The scatter in such relations can be tightened by weighting each EoS by the mass-radius posterior distributions on GW170817 \citep{ligo_scientific_collaboration_and_virgo_collaboration_properties_2019} and J0030+0451 \citep{miller_psr_2019} from the LIGO-Virgo Collaboration (LVC) and the Neutron Star Interior Composition Explorer (NICER), respectively. 
In addition, numerous studies have been carried out combining various combinations of multi-messenger NS observations and nuclear theory to constrain the EoS and predict the radius of a canonical $1.4 M_\odot$ NS, $R_{1.4}$. 
I thus compare two methods for predicting the baryonic masses of the three NSs considered here.
Method 1 one uses the baryonic-gravitational mass relation and Method 2 uses both the binary compactness relation and the B-C relation, the relation between the binding energy and compactness. 
\par
I use a set of 62 purely hadronic EoSs that are able to produce a NS above a maximum-mass lower limit of $1.97 M_\odot$ \citep{cromartie_relativistic_2020}. 
\begin{figure*}
\centering
\includegraphics[width=\textwidth]{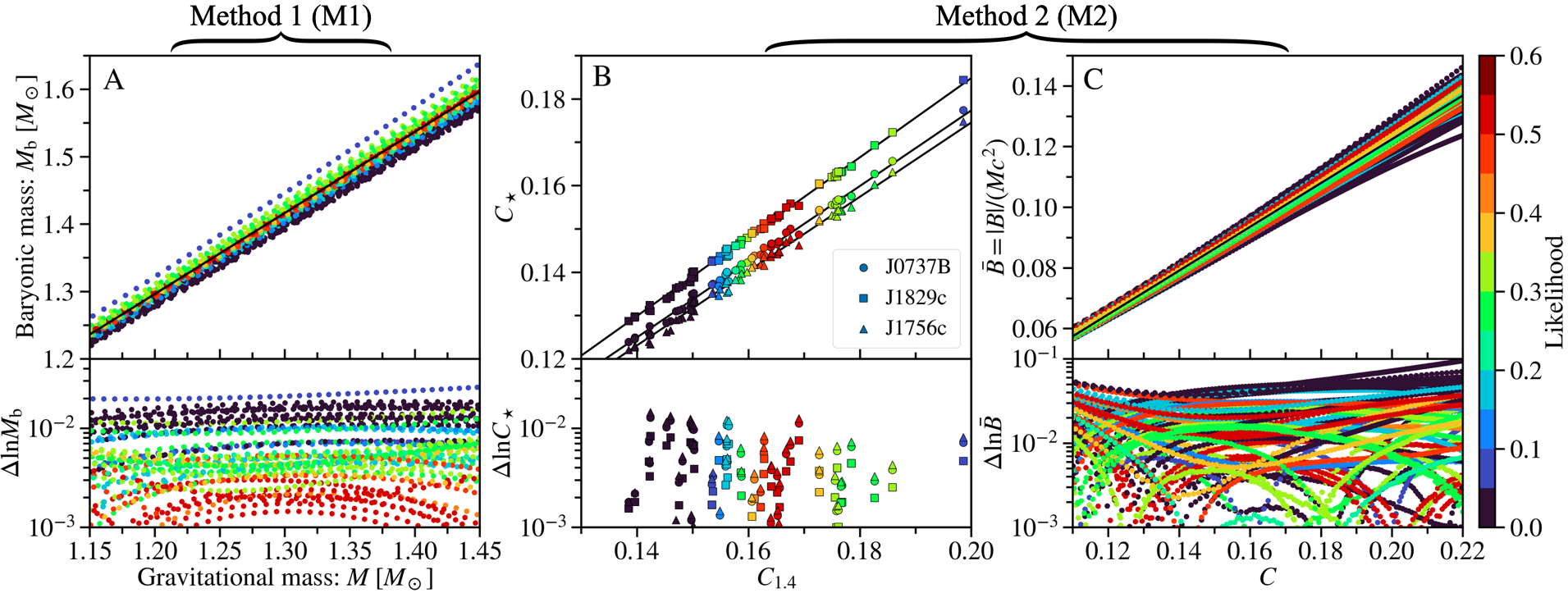}
\caption{\label{fig:fits} {\bf Weighted NS relations.} 
The top windows of each panel plots the data and fits for each relation. 
The color of each data point represents the likelihood of its respective EoS from both GW170817 and J0030+0451 (see Supplemental Material). 
The black lines correspond to the weighted best-fit lines for each relation. 
The bottom windows of each panel plots the residuals of the data points with respect to the best-fit line. 
Panel A: the baryonic-gravitational mass relation obtained over the range $M/M_\odot \in [1.15,1.45]$. 
Panel B: the binary compactness relation with $C_{\star} = (C_{1.2489}, C_{1.299}, C_{1.230})$. 
The circle, square, and triangle data points correspond to J0737B, J1829c, and J1756c, respectively. 
Panel C: the relation between the non-dimensional binding energy and compactness (B-C relation) obtained over the range $C \in [0.11,0.22]$. 
}
\end{figure*}
J0737B, J1829c, and J1756c are not observed to be spun-up to milli-second spin periods from accretion.
Thus, I model their interior structure in the non-rotating limit with the Tolman-Oppenheimer-Volkoff equations \citep{tolman_static_1939,oppenheimer_massive_1939}. 
The quantities I solve for as a function of the central mass density $\rho_{\rm c}$ include the gravitational mass $M$, radius $R$, baryonic mass $M_{\rm b}$, and nondimensional binding energy $\bar{B} = |B|/(M c^2)$, where $c$ is the speed of light.
Here, the NS binding energy is given by the difference between the gravitational and baryonic masses: $B = (M - M_{\rm b})c^2$.
The compactness $C = G M / (c^2 R)$ can also be obtained from the mass and radius, where $G$ is Newton's gravitational constant. 
\par
With this EoS catalog in hand, one can weight each EoS based on the radius it predicts at a gravitational mass of $1.4 M_\odot$ (see Supplemental Material). 
Both GW170817 and J0030+0451 have masses close to the canonical $1.4 M_\odot$ value. 
Thus, the radius posteriors on these NSs can be used to weight the EoS for the three low-mass NSs that are considered here.
This amounts to marginalizing the $M$-$R$ posteriors over the mass to get a distribution over the radius. 
The joint distribution is obtained from the GW170817 and J0030+0451 distributions by multiplying both distributions and normalizing. 
For a given EoS, the radius is fairly insensitive to changes in the gravitational mass for astrophysical low-mass NSs. 
\par
Here, I use weighted best-fit linear relations $y = a_0 + a_1 x$, where $y = (M_{\rm b}, C_\star, \bar{B})$, where the $\star$ subscript refers to the three NSs considered here, and $x = (M,C_{1.4},C)$, respectively. 
For the three NSs I consider, the inferred compactnesses are $C_{1.2489}$, $C_{1.299}$, and $C_{1.230}$ for J0737B, J1829c, and J1756c, respectively, where the numerical subscripts correspond to the median values of the measured gravitational mass.
The medians and uncertainties on the gravitational-mass are $1.2489 \pm 0.0007 M_\odot$, $1.299 \pm 0.0035 M_\odot$, and $1.230 \pm 0.007 M_\odot$, respectively. 
The EoS data and weighted fits for each relation are plotted in \autoref{fig:fits}. 
The coefficients and weighted average of the residuals with respect to this best-fit relation are tabulated in \autoref{tab:coeffs}.  
\begin{table}
\begin{tabular}{| l c c c c | c|}
   \hline
   \hline
   $y$ & $x$ & $a_0$ & $a_1$ & $\left\langle \Delta \ln y \right\rangle$ & $x \in [x_{\rm min}, x_{\rm max}]$\\
   \hline
   $M_{\rm b}$ & $M$ & $-0.147$ & $1.203$ & $0.39\%$ & $M\in [1.15,1.45]$\\ 
   $C_{1.2489}$ & $C_{1.4}$ & $2.674{\times}10^{-3}$ & $0.873$ & $0.39\%$ & -- \\ 
   $C_{1.299}$ & $C_{1.4}$ & $1.745{\times}10^{-3}$ & $0.915$ & $0.26\%$ & -- \\ 
   $C_{1.230}$ & $C_{1.4}$ & $3.034{\times}10^{-3}$ & $0.858$ & $0.44\%$ & -- \\ 
   $\bar{B}$ & $C$ & $-0.021$ & $0.718$ & $1.58\%$ & $C\in [0.11,0.22]$ \\ \hline
   \hline
\end{tabular}
\caption{The weighted linear relations used in this work, their corresponding numerical coefficients, weighted average residuals, and intervals over which they are defined. 
This relations are obtained with the catalog of 62 purely hadronic EoSs and weighted by the joint radius posteriors from GW170817 \citep{ligo_scientific_collaboration_and_virgo_collaboration_properties_2019} and J0030+0451 \citep{miller_psr_2019}.
}
\label{tab:coeffs}
\end{table}
As expected for the binary compactness relations, the weighted residual $\left\langle \Delta \ln C_\star \right\rangle$ tends towards zero as the mass approaches $1.4 M_\odot$, i.e., when the mass ratio approaches unity. 
\par
Given a precise gravitational-mass measurement, one can obtain the baryonic mass using the baryonic-gravitational mass relation (Method 1). 
This relation is obtained using a weighted linear fit to the EoS data over the gravitational-mass range $M/M_\odot \in [1.15,1.45]$. 
The data and the best fit are plotted in Panel A of \autoref{fig:fits}, where the weighted average of the residuals is $\left\langle \Delta \ln M_{\rm b}\right\rangle = 0.39\%$. 
For Method 1, the EoS uncertainty dominates the baryonic-mass estimate, which is informed by the mass-radius constraints from GW170817 and J0030+0451. 
\par
Method 2 is as follows: given the radius of a canonical $1.4 M_\odot$ NS from various previous studies, one can directly obtain $C_{1.4}$.
One can then obtain the compactness of our given NS of interest $C_\star$ by applying the binary compactness relation, where the approach is similar to the binary Love relation \citep{yagi_binary_2016}. 
One can then constrain the binding energy via the B-C relation, which is obtained over the range $C\in [0.11,0.22]$, which is relevant for the low-mass NSs considered here. 
The relations are plotted in Panels B and C of \autoref{fig:fits}. 
\par
For Method 2, the uncertainty on $R_{1.4}$ dominates the binding-energy estimate and thus the baryonic-mass estimate when given a gravitational-mass measurement at sub-percent-level precision. 
The $R_{1.4}$ constraints I use here come from Refs. \citep{capano_stringent_2020,essick_direct_2020,dietrich_multimessenger_2020,breschi_at2017gfo_2021,nicholl_tight_2021}, which I will refer to as C20, E20, D20, B21, and N21, respectively. 
These radius constraints are listed as follows: $R_{1.4} = \left(11.0_{-0.6}^{+0.9}, 12.54_{-0.63}^{+0.71}, 11.75_{-0.81}^{+0.86}, 12.2_{-0.8}^{+0.8}, 11.06_{-0.98}^{+1.01}\right)$ km at $90\%$ confidence, respectively. 
The consistency of these various studies suggests that multi-messenger NS observations are so far in agreement on the dense-matter EoS and resulting NS macroscopic properties \citep{al-mamun_combining_2021}. 
\par
\textbf{Testing the ECSN Scenario.} 
Using Methods 1 and 2, I obtain the baryonic mass and binding energy of J0737B, J1829c, and J1756c, which are thought to have formed from ultra-stripped progenitors. 
The main pieces of circumstantial evidence that suggest that these three NSs formed this way include their low gravitational masses and that their respective binary systems have low orbital eccentricities, $\lesssim 0.1$, and low transverse velocities \citep{dewi_formation_2004,piran_origin_2005,stairs_formation_2006,willems_formation_2006,ferdman_double_2013,ferdman_psr_2014,haniewicz_precise_2021}\footnote{Additional circumstantial evidence for J0737B's ECSN origin include Pulsar A's spin axis being closely aligned with the angular momentum of the orbital plane \citep{farr_spin_2011} and that Pulsar A's sense of rotation is prograde \citep{pol_direct_2018}. }.
\par
\begin{figure*}
\centering
\includegraphics[width=\textwidth]{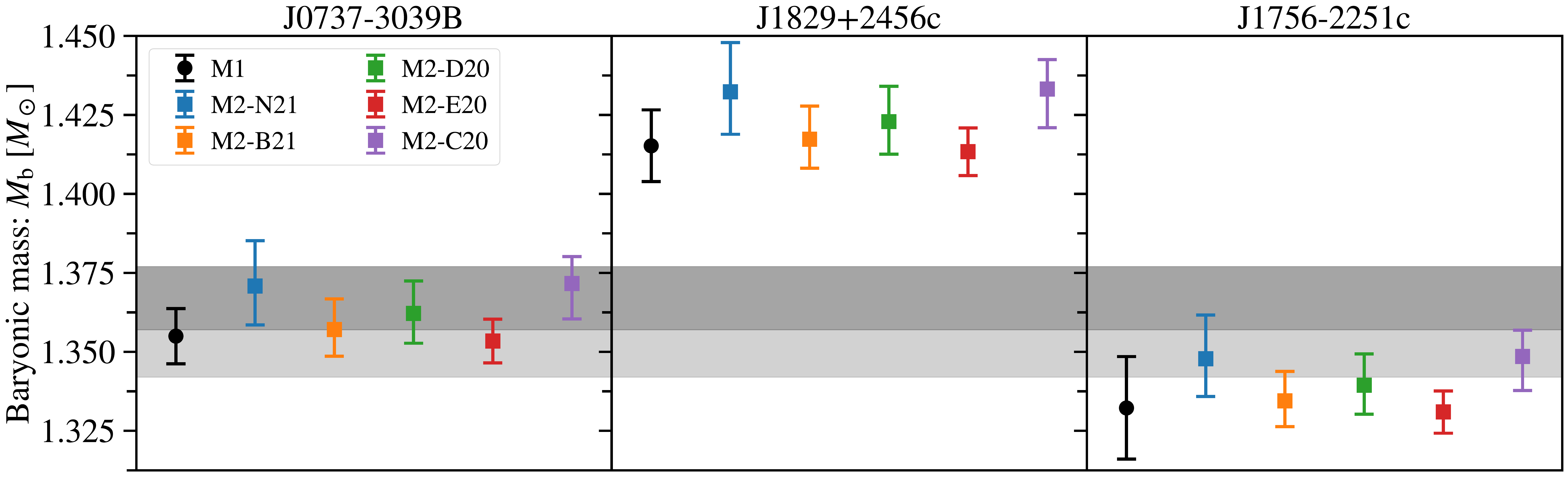}
\caption{\label{fig:bary} {\bf Baryonic-mass constraints.} 
The left, middle, and right panels correspond to J0737B, J1829c, and J1756c, respectively. 
For each NS, I compare the baryonic masses obtained from Methods 1 and 2, where the errorbars here are 90\% confidence intervals. 
The black circle errorbar corresponds to M1. 
The square errorbars correspond to M2, where each color uses a different $R_{1.4}$ constraint, with blue, orange, green, red, and purple colors corresponding to N21 \citep{nicholl_tight_2021}, B21 \citep{breschi_at2017gfo_2021}, D20 \citep{dietrich_multimessenger_2020}, E20 \citep{essick_direct_2020}, and C20 \citep{capano_stringent_2020}, respectively. 
The dark-grey shaded region corresponds to the ECSN mass range of 1.357-1.377$M_\odot$. 
The light-grey region corresponds to an allowed core mass loss of ${\le}0.015 M_\odot$. 
}
\end{figure*}

In \autoref{fig:bary}, I plot the baryonic-mass posteriors for J0737B, J1829c, and J1756c (left, middle, and right panels, respectively) along with the expected baryonic-mass range for ECSN and an allowed core mass loss of $\lesssim 0.015 M_\odot$ \citep{takahashi_evolution_2013,newton_testing_2018}. 
The constraints obtained with Methods 1 and 2 are statistically consistent with each other (within each other's 90\% confidence intervals), demonstrating the robustness of these different approaches. 
\par 
From the left panel of \autoref{fig:bary}, J0737B is consistent with forming via an ECSN with an ONeMg-core progenitor. 
The estimates that favor stiffer EoSs and thus less compact NSs overlap with the region with core mass loss $\le 0.015 M_\odot$. 
A low-mass iron core collapse that results in a weak SN kick may also explain J0737B's formation, thus motivating further explosion modeling to determine whether or not these two scenarios can be further distinguished. 
\par
From the middle panel of \autoref{fig:bary}, an ECSN origin for J1829c can be confidently disfavored.
Even with the estimates that favor softer EoSs, the discrepancy increases. 
J1829c thus significantly prefers an iron-core progenitor that results in a weak SN kick. 
This also motivates further modeling of ultra-stripped SNe with iron cores in order to characterize the core mass loss and resulting distribution of SN kicks. 
While the circumstantial evidence from the binary-system parameters favors a low-mass weak-kick scenario, adding in recent knowledge on the NS EoS has allowed for the ECSN scenario to be ruled out at high significance for J1829c. 
\par
From the right panel of \autoref{fig:bary}, the estimates slightly disfavor the ECSN hypothesis for J1756c, though it cannot yet be ruled out. 
At the very least, a core mass loss of $\Delta M_{\rm b} \sim 0.01$-$0.04 M_\odot$ is needed in order to reconcile the slight tension if J1756C did indeed form from an ECSN. 
The upper end of this mass-loss range is about a few times greater than what is expected based on numerical modeling. 
This also motivates resolved multi-dimensional modeling of ONeMg core collapse with detailed microphysics in order to further characterize the mass range for ECSNe and the core mass loss. 
\par
\textbf{Prospects for Constraining the EoS.}
Given the narrow mass-range for ECSNe, one might be able to constrain the EoS if one assumes a low-mass NS formed via this pathway \citep{podsiadlowski_double_2005}.
This also depends on how well our understanding of the explosion physics and net mass loss can be constrained. 
Further global modeling of ECSNe is needed in order to refine our understanding of the distribution of ONeMg baryonic core masses that will undergo an ECSN and what the outcome distribution of core mass loss is. 
Here, I consider J0737B, which has the smallest gravitational-mass uncertainty.
\par
\begin{figure}
\centering
\includegraphics[width=\columnwidth]{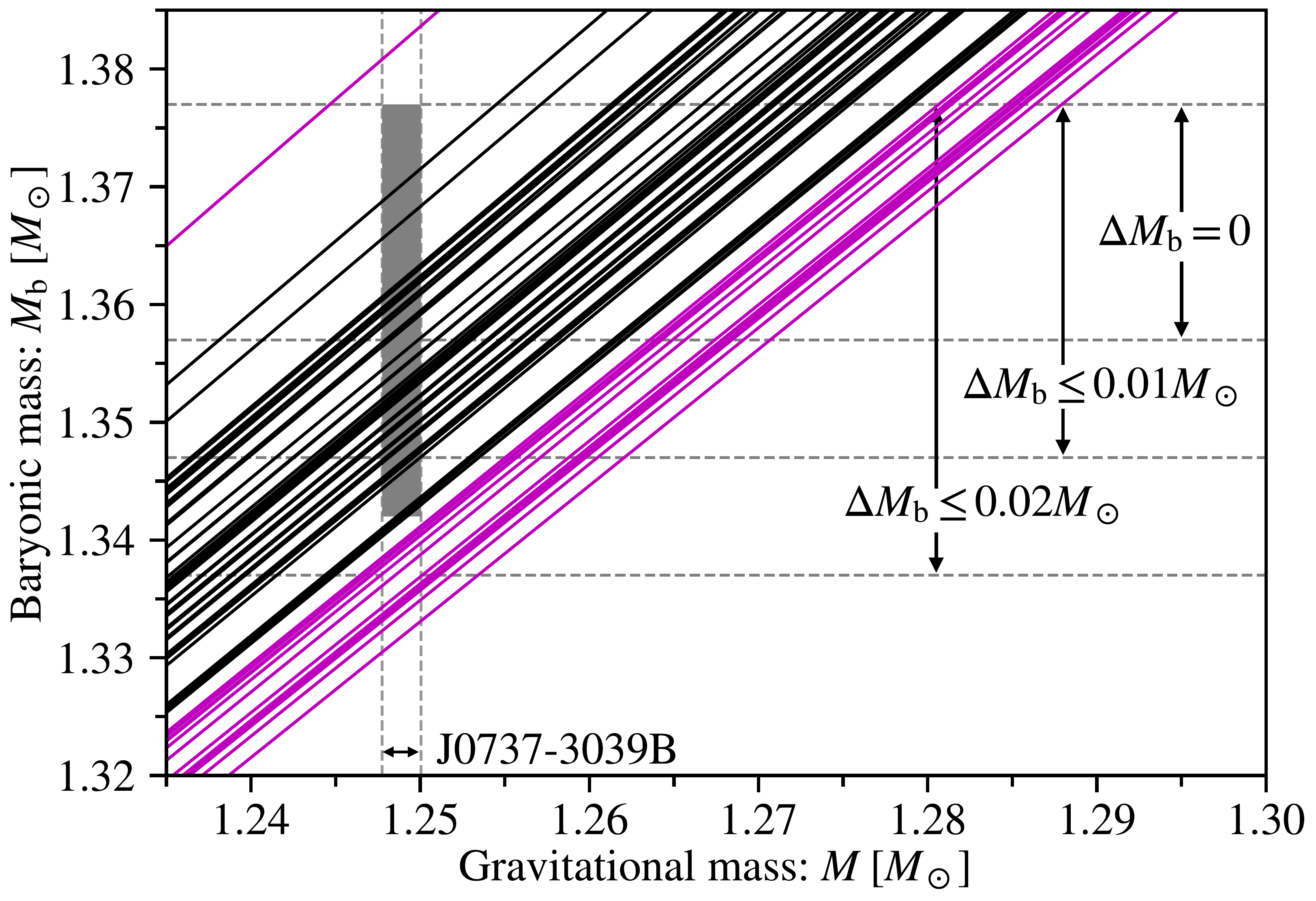}
\caption{\label{fig:eos} {\bf Prospects for constraining the EoS.} The baryonic mass vs. the gravitational mass from the EoS catalog. 
The dark-grey shaded box encompasses J0737B's gravitational mass uncertainty and the baryonic mass range of $1.357-1.377 M_\odot$ for ECSNe, along with an allowed core mass loss of $\Delta M_{\rm b} \le 0.015 M_\odot$. 
EoSs that fall within the dark-grey box are colored black, while those that fall outside are colored magenta and are thus disfavored, assuming J0737B indeed formed from an ECSN. 
}
\end{figure}

I plot in \autoref{fig:eos} the baryonic mass that each EoS in the catalog predicts as a function of the gravitational mass. 
The dark-grey shaded box encompasses the region of uncertainty for J0737B's mass as well as the baryonic mass range that is theoretically expected along with an allowed core mass loss of $\le 0.015 M_\odot$. 
Lines that fall within the gray box are colored black, while those that are outside are colored magenta. 
If the dark-grey box can be tightened from further ECSN modeling, then even more EoSs can be disfavored, assuming J0737B indeed formed from an ECSN. 
\par
\textbf{Conclusions.} 
In this work, I have obtained the first constraints on the baryonic mass and binding energy of three NSs thought to have formed from ultra-stripped progenitors, where the relations I used here incorporate both a purely hadronic EoS catalog and recent multi-messenger NS observations. 
I then used these constraints to test the formation scenarios of these NSs, where J0737B is consistent with having formed from an ONeMg core progenitor that underwent an ECSN, J1829c likely formed from an iron-core progenitor with a small kick, and J1756c slightly disfavors an ONeMg core progenitor, thought it cannot yet be ruled out. 
\par
These results further motivate continued improvements to the global modeling of ultra-stripped supernovae incorporating detailed microphysics and comparing iron core vs. ONeMg core progenitors \citep{suwa_neutrino-driven_2015,radice_electron-capture_2017,gessner_hydrodynamical_2018}.
Tightening the baryonic-mass range for ECSNe and the allowed core mass loss will allow for more stringent tests of these binary formation scenarios.  
Characterizing the distribution of kick velocities can be used as an additional way to test such supernova explosion models with DNS orbital parameters. 
Global modeling of the ejecta that remains bound within the remnant binary's Roche potential following the explosion may also be of interest to characterize any subsequent accretion of fallback material.
\par
Quantum chromodynamics may predict hyperonic and/or quarkyonic degrees of freedom at or above nuclear saturation density.
Strong first-order phase transitions, for example, result in a hybrid star with a smaller radius compared to a NS with the same gravitational mass and thus a larger binding energy. 
Additional NS measurements are thus still of merit to further constrain the dense-matter EoS as well as probe for the existence of exotic degrees of freedom and phase transitions. 
\par
Tighter EoS constraints from additional NS measurements will tighten the baryonic-mass constraints of the three NSs considered here and may confirm or rule out their respective ultra-stripped progenitors. 
The future also holds promise for new DNS systems that will be discovered with next-generation radio facilities like the Square Kilometer Array.
Such an observatory is expected to provide a larger sample of low-mass NSs with precisely measured gravitational masses to apply the methods that I present here and test their ultra-stripped progenitors. 
\par
\textbf{Acknowledgments.}~A.M.H. is supported by the McWilliams Postdoctoral Fellowship. 
I thank Paul Ricker and Alejandro C\'{a}rdenas-Avenda\~{n}o for detailed feedback on this manuscript. 
This work made use of the publicly available LVC posteriors on GW170817 \citep{ligo_scientific_collaboration_and_virgo_collaboration_properties_2019} and NICER Illinois-Maryland posteriors on J0030+0451 \citep{miller_psr_2019}.
\bibliography{references}
%
\newpage
\begin{center}
\bf{-- Supplemental Material --}
\end{center}
\textbf{Equation-of-State Catalog.} 
I use a set of 62 purely hadronic EoSs that are able to produce a NS with mass above a maximum-mass lower limit of $1.97 M_\odot$ \citep{cromartie_relativistic_2020}: ALF2, APR3, APR4, BCPM, BSk20, BSk21, BSk22, BSk23, BSk24, BSk25, BSk26, BSP, BSR2, BSR2Y, BSR6, BSR6Y, DD2, DD2Y, DDHd, DDME2, DDME2Y, ENG, FSU2, FSUGarnet, G3, GM1, GM1Y, GNH3, H4, IOPB, KDE0v1, MPA1, Model1, MS1, MS1b, NL3, NL3omegarho, NL3omegarhoY, NL3omegarhoYss, NL3Y, NL3Yss, Rs, SINPA, SK272, SK255, SKa, SKb, SkI2, SkI3, SkI4,  SkI5, SkI6, SKOp, SkMP, SLy, SLY230a, SLY2, SLY4, SLY9, TM1, WFF1, WFF2.
Further details on how these EoSs are generated can be found in Ref. \citep{silva_astrophysical_2020}. 
\par
\begin{figure}
\centering
\includegraphics[width=\columnwidth]{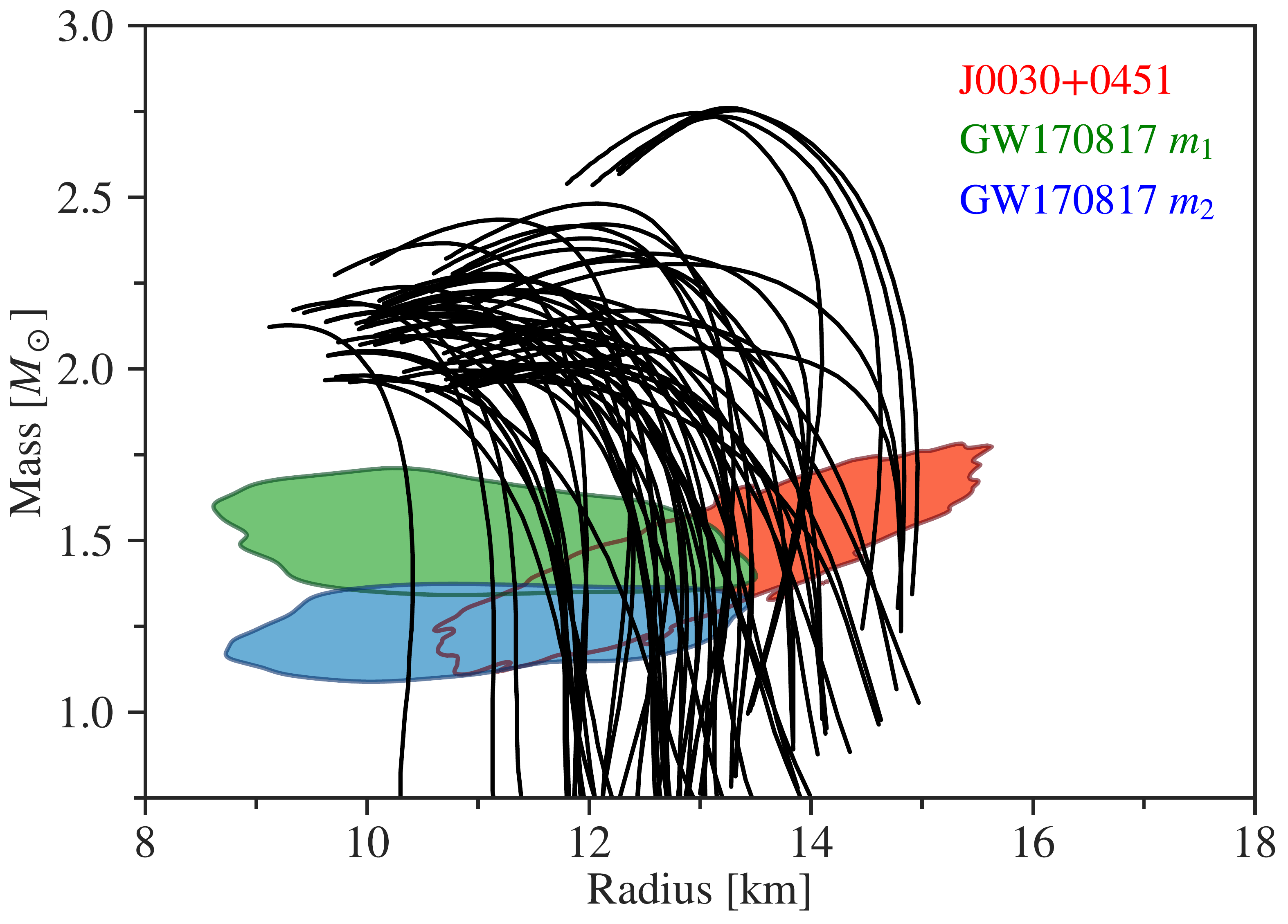}
\caption{\label{fig:mr} {\bf Mass-radius diagram.} The black curves are the $M$-$R$ curves predicted from each of the 62 EoSs in the catalog used here.
The green, blue, and red shaded regions are the posterior distributions at $90\%$ confidence for GW170817's primary, GW170817's secondary, and J0030+0451. 
}
\end{figure}
\begin{figure}
\centering
\includegraphics[width=\columnwidth]{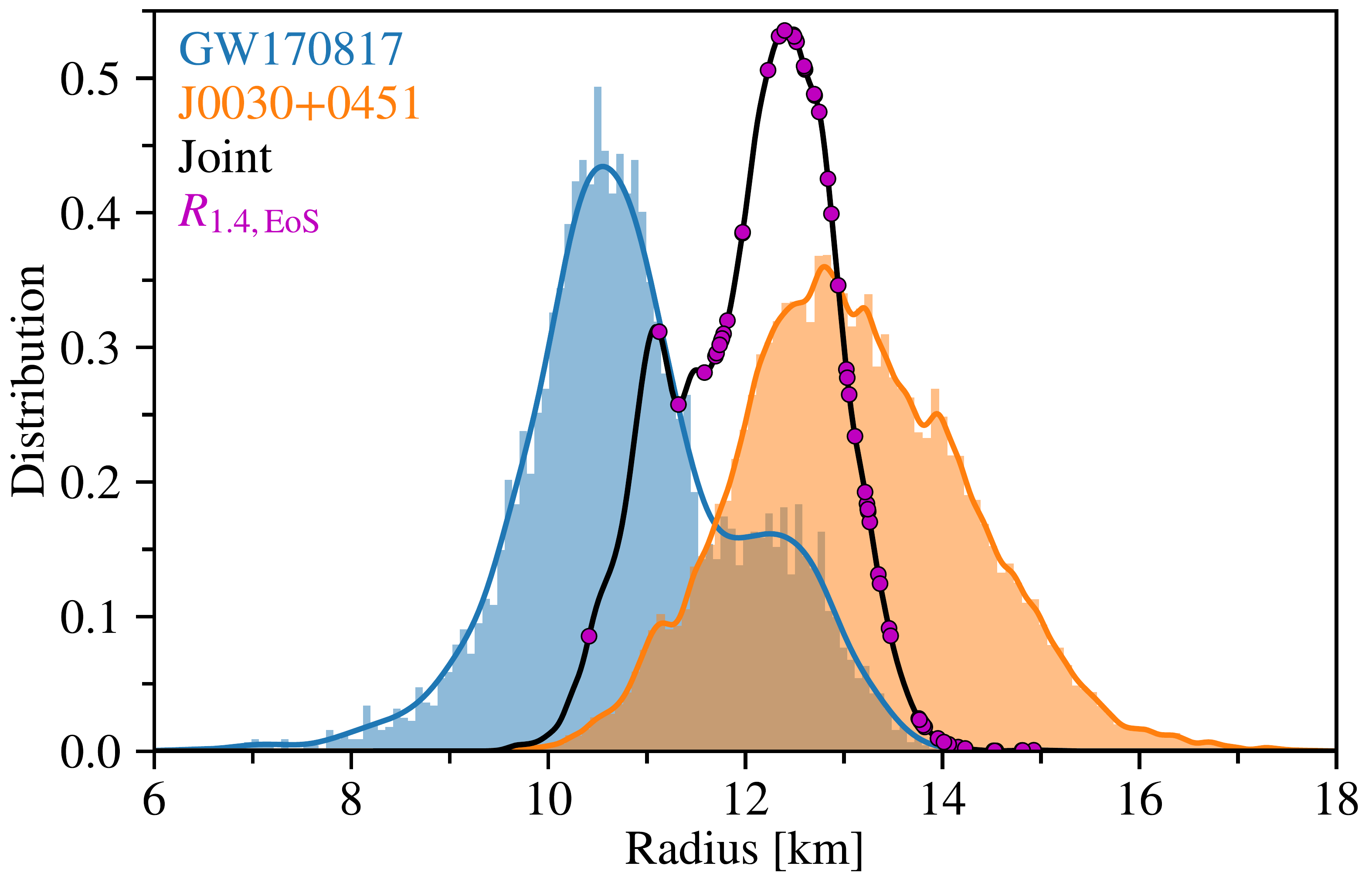}
\caption{\label{fig:joint} {\bf Joint constraints on the NS radius.} The blue and orange histograms and their kernel-density estimates correspond to the radius distributions from GW170817 (both the primary and secondary) and J0030+0451, respectively. 
The black distribution corresponds to the joint distribution, which is obtained by multiplying the blue and orange distributions and normalizing. 
The magenta points overlayed on the joint distribution correspond to each of the 62 EoSs in the catalog, where their $x$ position is obtained from their respective prediction for $R_{1.4}$. 
}
\end{figure}

The mass-radius curve for each EoS is plotted in \autoref{fig:mr}. 
The joint mass-radius posteriors for GW170817 and J0030+0451 at $90\%$ confidence are also plotted as the green, blue, and red shaded regions. 
Here, I consider the full radius posteriors, i.e., marginalizing over the gravitational mass, and I plot the resulting distributions for GW170817 and J0030+0451 in \autoref{fig:joint}. 
The joint distribution is obtained by multiplying the two distributions and normalizing. 
The weight used for each EoS in the weighted linear fits is given by the likelihood from the joint distribution. 
\par
\textbf{Weighted NS Relations.}
With this EoS catalog, I solve the Tolman-Oppenheimer-Volkoff equations for $M$, $R$, $M_{\rm b}$, and $\bar{B}$ as functions of the central density $\rho_{\rm c}$. 
The NS's gravitational mass is given by
\begin{equation}
M = 4 \pi \int_0^R \epsilon (r) r^2 \, {\rm d}r \ ,
\end{equation}
where $r$ is the radial coordinate and $\epsilon$ is the mass-energy density of the NS fluid. 
The NS's baryonic mass is
\begin{multline}
M_{\rm b} = 4 \pi m_{\rm b} \int_0^R n (r) r^2 \left(1 - \frac{2 m(r)}{r}\right)^{-1/2} {\rm d}r \\ = 4 \pi \int_0^R \rho (r) r^2 \left(1 - \frac{2 m(r)}{r}\right)^{-1/2} {\rm d}r \ ,
\end{multline}
where $m_{\rm b}$ is the mass of the baryon, $n$ is the baryon number density, $m$ is the gravitational mass enclosed within $r$, and $\rho$ is the baryon mass density. 
Given a gravitational-mass posterior, I obtain the posterior on the baryonic mass as
\begin{equation}
{\cal P}( M_{\rm b}) = \int {\cal P} (M_{\rm b} | M) {\cal P} (M) \, {\rm d}M \ ,
\end{equation}
where I use the weighted average of the residuals as the standard deviation of a Gaussian distribution
\begin{equation} \label{eq:gaussian}
{\cal P} (M_{\rm b} | M) = \left(2 \pi \sigma^2\right)^{-1/2} \exp \left[-(M_{\rm b} - M_{\rm b,fit})/(2 \sigma^2)\right] \ ,
\end{equation}
where $\sigma = \left\langle \Delta \ln M_{\rm b} \right\rangle$. 
This approach is similar to how NS properties are inferred in Ref. \cite{kumar_inferring_2019}.  
The baryonic mass and binding energy I obtain with the baryonic-gravitational mass relation (Method 1) along with their $1\sigma$ uncertainties are tabulated in \autoref{tab:masses}. 
\begin{table}
\begin{tabular}{ l c c c}
   \hline \hline
   Quantity  & J0737-3039B & J1756-2251c & J1829+2456c \\
   \hline
   $M/M_\odot$ & $1.2489 \pm 0.0007$ & $1.230 \pm 0.007$ & $1.299 \pm 0.0035$ \\ 
   $M_{\rm b}/M_\odot$ & $1.3555\pm 0.0053$ & $1.332 \pm 0.010$ & $1.4152 \pm 0.0069$ \\ 
   $|B|/(M_\odot c^2)$ & $0.1061 \pm 0.0053$ & $0.102 \pm 0.012$ & $0.1162\pm 0.0077$ \\ \hline
   \hline
\end{tabular}
\caption{NSs that may have formed from an ultra-stripped SN, their gravitational masses measured from radio timing, their inferred bayronic masses using the baryonic-gravitational mass relation (Approach 1), and their binding energies. 
The errorbars correspond to the $1\sigma$ uncertainties. 
}
\label{tab:masses}
\end{table}
\par
Another approach is to use recently reported multi-messenger constraints on the radius of a $1.4 M_\odot$ NS and infer binding energy of a given NS of interest using the binary compactness relation and the B-C relation. 
First consider the binary compactness relation. 
Given that NS matter is thought to be described with a single EoS, the macroscopic properties of two slowly rotating NSs may be similar if they have similar gravitational masses, i.e., a mass ratio near unity. 
The binding energy can then be obtained by relating the non-dimensional binding energy ${\bar B}$ with the compactness. 
Given a constraint on $R_{1.4}$, one can then obtain $\bar{B}$ using the binary compactness relation and the B-C relation as follows
\begin{equation}
{\cal P}(\bar{B}) = \int \int {\cal P}(\bar{B}|C_\star) \cdot {\cal P}(C_\star | C_{1.4}) \cdot {\cal P} (C_{1.4}) \ {\rm d} C_\star {\rm d} C_{1.4} \ ,
\end{equation}
where the distributions ${\cal P}(\bar{B} | C_\star)$ and ${\cal P} (C_\star | C_{1.4})$ have the same form as \autoref{eq:gaussian} and use their respective weighted residuals $\langle \Delta \ln \bar{B} \rangle$ and $\left\langle \Delta \ln C_\star \right\rangle$. 
The baryonic mass can then be obtained by adding the gravitational mass with the absolute value of the binding energy. 
%
\end{document}